\documentclass[aps,prl,twocolumn]{revtex4}
\usepackage{amsfonts}
\usepackage{color}
\usepackage{eepic}
\usepackage{epic}
\usepackage{graphicx}
\begin{document}
\preprint{}
\title{A Complete Characterization of Mixed State Entanglement using Probability Density Functions}

\author{Shanthanu Bhardwaj} 
\email{shanth@iitk.ac.in}
\author{V. Ravishankar}
\email{vravi@iitk.ac.in}
\affiliation{Department of Physics, Indian Institute of Technology, Kanpur-208016, INDIA}
\date{\today}
\begin{abstract}
We propose that the entanglement of mixed  states is characterised 
properly in terms of a probability density function $\mathcal{P}(\mathcal{E})$.  There is a need for such a measure since the prevalent measures (such as \textit{concurrence} and \textit{negativity}) are rough benchmarks, and not monotones of each other. Considering the specific case of two qubit mixed states, we provide an explicit construction
of $\mathcal{P}(\mathcal{E})$  and show that it is characterised by a  set of
parameters, of which concurrence is but one particular combination. $\mathcal{P}(\mathcal{E})$ is manifestly invariant under  $SU(2) \times SU(2)$
transformations.  It can, in fact,  reconstruct the state up to
local operations - with the specification of at most 
four additional parameters.  Finally the new measure  resolves the controversy regarding the role of entanglement in quantum computation in NMR systems.
\end{abstract}
\pacs{}
\keywords{}
\maketitle

Quantum entanglement   is a 
 unique resource for novel (nonclassical) applications such as quantum algorithms
\cite{Shor:1994jg+Grover:1996rk}, quantum cryptography \cite{BBBSS+BB84}, and more
recently, metrology \cite{nature}. Thus, it has a pivotal role in quantum information
theory.  It is also central to the study of the foundations of 
quantum mechanics \cite{Bell:1964kc+CHSH}. 
However, while  pure state entanglement  is well defined, 
mixed state entanglement  (MSE) is still rather poorly understood.
 Currently used definitions  such as entanglement of formation (EOF)  \cite{Wootters:1997id} and 
 separability \cite{Werner} are based on the emphasis given to
   a particular quantum feature.
  These definitions are not  equivalent \cite{Grudka} and are operational in a 
  limited sense. 
 Thus, concurrence
 as a characteristic of EOF \cite{Hill:1997pf} is defined only for a two qubit system; 
  negativity as a 
 criterion for separability \cite{Peres:1996dw, Horodecki:1997vt} is necessary and 
 sufficient only for two qubit systems 
 and a qubit-qutrit system. Likewise, 
 majorization \cite{nielsen} is  a necessary condition for separability. Further, 
 concurrence and negativity are not relative monotones, although the former bounds the
 latter from above. In particular, states with the same negativity 
 may have different concurrence and vice versa.
 Note that  real systems are almost always in a mixed state. Indeed, 
 NMR quantum computers (NMR QC) \cite{Gershenfeld, Cory} are prepared in the so 
 called pseudo pure states
 which are highly mixed. Their concurrence (and hence negativity) is zero, and yet
 nontrivial nonclassical gate operations with up to eight qubits have been reported
 \cite{Anil}. More recently, a 12-qubit pseudopure state has been reported for a weakly coupled NMR system \cite{12-qubit}.
  To unravel the sense in which the entanglement is a resource in these systems,
 there is a clear need to go beyond the above mentioned benchmarks.
 We address this problem here, and propose
  an alternative definition of MSE.
  
  To motivate our approach, we recall that a mixed state is required to describe an
  ensemble of quantum systems each of which is in a pure state.
  Entanglement has a sharp value in each pure state; Thus,
   MSE may be expected to acquire a statistical character, and be characterized 
   by a suitably
  defined probability density function (PDF). We propose below a definition of MSE, in terms of
  one such  PDF, which is strictly operational and applicable to any bipartite system.
   The definition
  \textit{does not} require any new notion of entanglement other than that for pure states.
   We proceed to give an explicit construction of the PDF for the important case
   of two qubit systems. For these sytems, 
    we show that the PDF has some striking morphological features which completely
   encode the information on MSE: these features appear as  a few points of 
   discontinuity of various orders in the PDF. These points are shown to allow an almost
   complete reconstruction of the state, up to local operations (LO).
   It is shown how concurrence gets reinterpreted as
   a benchmark. Finally, the issue of entanglement in NMR QC gets naturally resolved.

We now posit a probability density function for entanglement, $ \mathcal{P}_{\rho}(\mathcal{E})$.
The definition will be given in several steps:
Let the state $\rho$ of a  two qubit system   be characterised by its eigenvalues 
$\lambda^{\downarrow}_i$, with 
respective eigenstates $\vert \psi_i \rangle$ which are orthonormal;  the notation implies
that the eigenvalues are arranged in a non increasing order. 
The choice of $\vert \psi_i \rangle$ is non unique if the 
eigenvalues are degenerate, but it is of no concern to us here. 
(i) As the first step,
we define
a sequence of projection operators  $\Pi_i = \displaystyle\sum_{j=1}^{i} 
 |\psi_j\rangle \langle \psi_j|$; $\Pi_i \subset \Pi_{i+1}$, with
$ \Pi_4$ being the full Hilbert space. It is a trivial identity that
 \begin{eqnarray}
 \nonumber \rho &=& (\lambda_1 - \lambda_2)\Pi_1 +  (\lambda_2 - \lambda_3)\Pi_2 + \\
 & & (\lambda_3 - \lambda_4)\Pi_3 +  \lambda_4\Pi_4.
 \end{eqnarray}
 The above equation resolves $\rho$ into an incoherent sum of a hierarchy of the subspaces $\Pi_i$,
 with the weights given by the nonnegative vector $ \Lambda= (\lambda_1 - \lambda_2,~ 
 \lambda_2 - \lambda_3,~ \lambda_3 - \lambda_4,~ \lambda_4)$,  whose norm is a 
 measure of the purity of the state. In a sense, the vector represents the manner in which
the state ``spills over'' to successively higher dimensional spaces. (ii) As the next step,  observe that if $\rho$ is a projection
 $\Pi_i$ of dimension $i$, the ensemble would  be uniformly distributed over states in $\Pi_i$: $\langle \psi \vert \rho
\vert \psi \rangle = 1 \forall |\psi \rangle \in \Pi_i$. 
 A Probability Density Function (PDF) may be naturally defined thus:
 \begin{eqnarray}
 \mathcal{P}_{i}(\mathcal{E}) = \frac{\int \int d\mathcal{E'} d\mathcal{H}_i \, 
\delta(\mathcal{E'} - \mathcal{E})}{\int d\mathcal{H}_i}
\end{eqnarray}
 with $d\mathcal{H}_i$ being the appropriate Haar measure. (iii) As the last step, rescale $\rho \rightarrow \rho_s$ and 
 rewrite it in terms of the difference in relative weights $\mu_i=(\lambda_i -
 \lambda_{i+1})/\lambda_1$ as
 $\rho_s = \sum_i^4 \mu_i \Pi_i$. The definition of the PDF is then given by the simple superposition
 \begin{eqnarray}
 \mathcal{P}_{\rho}(\mathcal{E}) = \sum_i^4 \mu_i\mathcal{P}_{i}(\mathcal{E}).
 \end{eqnarray}
The rest of the paper is devoted to an elucidation of Eqns. 2 and 3. We choose the pure state concurrence
$ 2 \vert \alpha_{\uparrow \uparrow}
\alpha_{\downarrow \downarrow} - 
\alpha_{\uparrow \downarrow} \alpha_{\downarrow \uparrow} \vert$,
in terms of the coefficients of expansion of $|\psi \rangle = \alpha_{\uparrow \uparrow} |\uparrow\uparrow\rangle + \alpha_{\uparrow \downarrow} |\uparrow\downarrow\rangle + \alpha_{\downarrow \uparrow} |\downarrow\uparrow\rangle + \alpha_{\downarrow \downarrow} |\downarrow\downarrow\rangle$, as the measure of pure state entanglement.
Although the definition is given for the simplest case, the generalization to higher spin systems is straightforward, 
 and we do not discuss it any further in this paper. 
 
\section{Two qubit probability density functions: Description of Subspaces}
We first consider the situation when $\rho$ is a projection, case by case. 
We then move on
to discuss the general case (displayed in Eqn.3). Since the normalization is provided
by dividing by the total volume of the group space,
 the trace factors will be dropped. We employ LO on the subspaces
freely, since the PDF remains unaffected.

\noindent{\bf The pure state}: Consider $\rho=\Pi_1 \equiv \vert \phi \rangle \langle \phi \vert$. 
The probability density function $\mathcal{P}_{1}(\mathcal{E})$ is simply $\delta(\mathcal{E} - \mathcal{E}_{\phi})$, in terms of the
the entanglement of $\vert \phi\rangle$.  
The PDF is singular, and specified by a single number.

\noindent{\bf Two dimensional projection}: $\rho=\Pi_2$ is the most complicated and the most interesting case. 
Suppose $\vert \psi \rangle \in \Pi_2$. Let $\vert \chi_1\rangle,~ \vert\chi_2\rangle$ 
be orthonormal and span $\Pi_2$. We have, $ \vert \psi \rangle = 
\vert \chi_1 \rangle \cos{\frac{\theta}{2}} e^{i\phi / 2} + 
\vert \chi_2\rangle \sin{\frac{\theta}{2}} e^{- i\phi / 2}$. 
The Haar measure is simply read off as
$d\mathcal{H} = \sin \theta d\theta d\phi$. By a suitable LO,  
 we can choose $\vert \chi_1 \rangle$ to be separable, 
in its canonical
form $(1,0,0,0)$ in a separable basis, \textit{i.e.}, 
$\vert \chi_1 \rangle = \vert \uparrow \uparrow \rangle$. 
 $\vert \chi_2 \rangle$ 
can be further
chosen to be of the form $(0, x, y, +\sqrt{1-x^2-y^2})$, where $x,y \ge 0$. 
The entanglement
distribution is, therefore, characterized by two non-negative parameters, 
and is implicitly
determined by Eqn.2. 

\begin{figure}[ht]
\setlength{\unitlength}{0.120450pt}
\begin{picture}(1800,1500)(0,0)
\footnotesize
\color{black}
\thicklines \path(370,249)(411,249)
\thicklines \path(1676,249)(1635,249)
\put(329,249){\makebox(0,0)[r]{$0$}}
\color{black}
\thicklines \path(370,483)(411,483)
\put(329,483){\makebox(0,0)[r]{$2$}}
\color{black}
\thicklines \path(370,717)(411,717)
\put(329,717){\makebox(0,0)[r]{$4$}}
\color{black}
\thicklines \path(370,950)(411,950)
\put(329,950){\makebox(0,0)[r]{$6$}}
\color{black}
\thicklines \path(370,1184)(411,1184)
\put(329,1184){\makebox(0,0)[r]{$8$}}
\color{black}
\thicklines \path(370,1418)(411,1418)
\put(329,1418){\makebox(0,0)[r]{$10$}}
\color{black}
\thicklines \path(370,249)(370,290)
\thicklines \path(370,1418)(370,1377)
\put(370,166){\makebox(0,0){$0$}}
\color{black}
\thicklines \path(631,249)(631,290)
\thicklines \path(631,1418)(631,1377)
\put(631,166){\makebox(0,0){$0.2$}}
\color{black}
\thicklines \path(892,249)(892,290)
\thicklines \path(892,1418)(892,1377)
\put(892,166){\makebox(0,0){$0.4$}}
\color{black}
\thicklines \path(1154,249)(1154,290)
\thicklines \path(1154,1418)(1154,1377)
\put(1154,166){\makebox(0,0){$0.6$}}
\color{black}
\thicklines \path(1415,249)(1415,290)
\thicklines \path(1415,1418)(1415,1377)
\put(1415,166){\makebox(0,0){$0.8$}}
\color{black}
\thicklines \path(1676,249)(1676,290)
\thicklines \path(1676,1418)(1676,1377)
\put(1676,166){\makebox(0,0){$1$}}
\color{black}
\color{black}
\thicklines \path(370,249)(1676,249)(1676,1418)(370,1418)(370,249)
\color{black}
\put(82,833){\makebox(0,0)[l]{\shortstack{$\mathcal{P(E)}$}}}
\color{black}
\put(1023,42){\makebox(0,0){$\mathcal{E}$}}
\color{magenta}
\thicklines 
\dottedline[*]{30}(370,405)(1350,405)(1350,249)
\color{blue}
\thicklines 
\dottedline[x]{50}(370,249)(410,256)(449,263)(489,270)(528,278)(568,285)(607,292)(647,300)(687,307)(726,315)(766,323)(805,332)(845,340)(884,349)(924,359)(964,368)(1003,379)(1016,382)

\dottedline[x]{50}(1016,382)(1056,393)(1096,405)(1135,418)(1175,432)(1214,447)(1254,464)(1293,483)(1333,504)(1373,529)(1412,559)(1452,595)(1491,641)(1531,703)(1570,795)(1610,956)(1636,1177)(1650,1394)

\color{red}
\thicklines \path(370,249)(370,249)(370,249)(370,249)(383,249)(396,249)(409,250)(422,251)(448,252)(461,255)(501,259)(514,262)(527,265)(540,266)(566,268)(579,269)(592,271)(605,274)(618,277)(631,279)(644,280)(657,281)(670,283)(683,286)(697,291)(710,296)(723,294)(736,300)(749,299)(762,302)(775,304)(788,306)(801,307)(814,310)(827,312)(840,315)(853,316)(866,319)(879,320)(892,323)(905,326)(919,328)(932,330)(945,334)(958,335)(971,339)(984,340)(997,345)(1010,347)(1023,350)(1036,354)
\thicklines \path(1036,354)(1049,357)(1062,360)(1075,364)(1088,367)(1101,371)(1114,375)(1127,379)(1141,384)(1154,388)(1167,393)(1180,398)(1193,403)(1206,410)(1219,416)(1232,422)(1245,429)(1258,437)(1271,446)(1284,455)(1297,465)(1310,477)(1323,491)(1336,506)(1350,524)(1363,547)(1376,575)(1389,615)(1402,682)(1415,853)(1428,726)(1441,659)(1454,624)(1467,602)(1480,585)(1493,572)(1506,561)(1519,552)(1532,544)(1532,249)(1545,249)(1558,249)(1572,249)(1585,249)(1598,249)(1611,249)(1624,249)(1637,249)(1650,249)(1663,249)

\color{black}
\thicklines \path(370,249)(1676,249)(1676,1418)(370,1418)(370,249)
\end{picture}

\caption{Some Typical probability density functions for $\Pi_2$.  Note the solid  curve,
 which shows all the features of $\mathcal{P}_{2}(\mathcal{E})$.  
 It has a cusp at $\mathcal{E}_{cusp} = 0.8$ and goes to zero at 
 $\mathcal{E}_{max} = 0.89$.  
 The step function is an extreme example, where $\mathcal{E}_{cusp} = 0$, 
and the other dotted curve, has $\mathcal{E}_{cusp} = \mathcal{E}_{max} = 1$}
\end{figure}
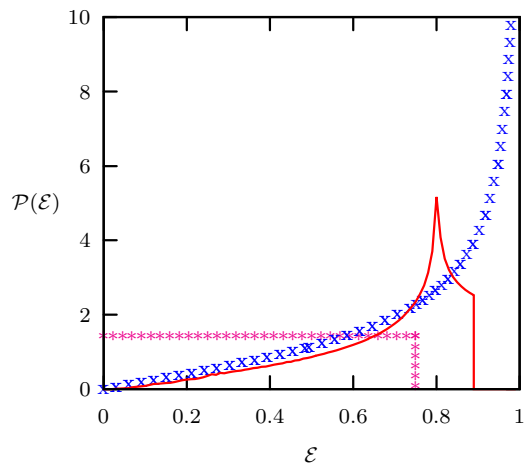
The generic form of the PDF in $\Pi_2$ is shown in FIG. 1 (the solid curve).
We observe that it has three markers,  (i) $\mathcal{E}_{cusp}$, the entanglement at which the probability density diverges, 
invariably as 
a cusp, (ii)$\mathcal{E}_{max}$, the maximum entanglement allowed, and 
(iii) $\mathcal{P}_2(\mathcal{E}_{max})$,  the probability density at $\mathcal{E}_{max}$. 
In fact, any two of them suffice to characterise the PDF completely.
 One may  specify {\it e.g.},
$(\mathcal{E}_{max},~ \mathcal{P}_2(\mathcal{E}_{max}))$, or equivalently,
$(\mathcal{E}_{cusp},~ \mathcal{P}_2(\mathcal{E}_{max}))$ for characterizing the curve. A straightforward
computation establishes the relations

 \begin{eqnarray}
\mathcal{E}_{max} &=& xy + \sqrt{z^2 + x^2y^2} \\
 \mathcal{E}_{cusp} &=& \frac{z^2}{\mathcal{E}_{max}} = \mathcal{E}_{max} \cos \mu \\
 \nonumber \mu &=& \sin^{-1}\left(\frac{1}{\mathcal{E}_{max}\mathcal{P}_2(\mathcal{E}_{max})}\right) \\
  &=& \sin^{-1}\left( \frac{2 \sqrt{xy(xy\mathcal{E}_{max} + z^2)}}
 {\mathcal{E}_{max}^{3/2}} \right) 
\end{eqnarray}
which allow us to determine the parameters $x,y$ that define $\Pi_2$. $\mu$ is well defined
by virtue of the inequality,
$\mathcal{P}_2(\mathcal{E}_{max}) \geq  1/\mathcal{E}_{max}$.
Note that unlike with the other measures, the state itself can be  reconstructed up to LO.

Two extreme cases occur when $\mathcal{E}_{cusp} =0$ and $\mathcal{E}_{cusp} =\mathcal{E}_{max}$. 
In the first case,
the PDF is a step function, terminating at some $\mathcal{E}_{max}$.
 In the second case, the density increases monotonically, diverging at $\mathcal{E}_{max}$
(FIG. 1).
The relative abundance of entangled states is more in the latter case. 
One may \textit{per se}
expect that the associated concurrence should also be larger. Interestingly, however,  the concurrence 
is related to the new parameters by
$\mathcal{C}=(\mathcal{E}_{max} - \mathcal{E}_{cusp})/2$, vanishing when 
$\mathcal{E}_{cusp}=\mathcal{E}_{max}$. In other words,  it is not sensitive to the relative 
abundance at zero (or small entanglements) at all. 
 In any case,
$\mathcal{C}$ emerges as a particular benchmark of the probability density, characterizing it
only partially. We note that if $\rho = \Pi_3 ~~\rm{or}~~ \Pi_4$, then its concurrence is zero.
By the convexity of the concurrence, we conclude that 
 the  concurrence $\mathcal{C}_{\rho}$ is bounded by 
$$
\mathcal{C}_{\rho} \leq (\lambda_1 - \lambda_2) \mathcal{C}_{\Pi_1} + 
(\lambda_2 - \lambda_3) \mathcal{C}_{\Pi_2}. 
$$
Incidentally, the entanglement distribution of a subspace $\Pi_2^{c}$ orthogonal to 
$\Pi_2$ is the
same as that of $\Pi_2$.

\noindent{\bf Three dimensional projection}: We now move on to the case $\rho= \Pi_3$, whose PDF has a simpler structure.  $\Pi_3$ is
completely characterised by its dual, $\vert \xi \rangle \perp \Pi_3$. Thus,
 the PDF is characterised by a single parameter $\mathcal{E}_{\perp}$, which is the entanglement of the orthogonal state $\vert \xi \rangle$.

\begin{figure}[ht]


\definecolor{grey}{rgb}{.2, 0.2, .2}
\setlength{\unitlength}{0.120450pt}
\begin{picture}(1800,1500)(0,0)
\footnotesize
\color{black}
\thicklines \path(370,249)(411,249)
\thicklines \path(1676,249)(1635,249)
\put(329,249){\makebox(0,0)[r]{ 0}}
\color{black}
\thicklines \path(370,379)(411,379)
\thicklines \path(1676,379)(1635,379)
\put(329,379){\makebox(0,0)[r]{ 0.2}}
\color{black}
\thicklines \path(370,509)(411,509)
\thicklines \path(1676,509)(1635,509)
\put(329,509){\makebox(0,0)[r]{ 0.4}}
\color{black}
\thicklines \path(370,639)(411,639)
\thicklines \path(1676,639)(1635,639)
\put(329,639){\makebox(0,0)[r]{ 0.6}}
\color{black}
\thicklines \path(370,769)(411,769)
\thicklines \path(1676,769)(1635,769)
\put(329,769){\makebox(0,0)[r]{ 0.8}}
\color{black}
\thicklines \path(370,898)(411,898)
\thicklines \path(1676,898)(1635,898)
\put(329,898){\makebox(0,0)[r]{ 1}}
\color{black}
\thicklines \path(370,1028)(411,1028)
\thicklines \path(1676,1028)(1635,1028)
\put(329,1028){\makebox(0,0)[r]{ 1.2}}
\color{black}
\thicklines \path(370,1158)(411,1158)
\thicklines \path(1676,1158)(1635,1158)
\put(329,1158){\makebox(0,0)[r]{ 1.4}}
\color{black}
\thicklines \path(370,1288)(411,1288)
\thicklines \path(1676,1288)(1635,1288)
\put(329,1288){\makebox(0,0)[r]{ 1.6}}
\color{black}
\thicklines \path(370,1418)(411,1418)
\thicklines \path(1676,1418)(1635,1418)
\put(329,1418){\makebox(0,0)[r]{ 1.8}}
\color{black}
\thicklines \path(370,249)(370,290)
\thicklines \path(370,1418)(370,1377)
\put(370,166){\makebox(0,0){ 0}}
\color{black}
\thicklines \path(631,249)(631,290)
\thicklines \path(631,1418)(631,1377)
\put(631,166){\makebox(0,0){ 0.2}}
\color{black}
\thicklines \path(892,249)(892,290)
\thicklines \path(892,1418)(892,1377)
\put(892,166){\makebox(0,0){ 0.4}}
\color{black}
\thicklines \path(1154,249)(1154,290)
\thicklines \path(1154,1418)(1154,1377)
\put(1154,166){\makebox(0,0){ 0.6}}
\color{black}
\thicklines \path(1415,249)(1415,290)
\thicklines \path(1415,1418)(1415,1377)
\put(1415,166){\makebox(0,0){ 0.8}}
\color{black}
\thicklines \path(1676,249)(1676,290)
\thicklines \path(1676,1418)(1676,1377)
\put(1676,166){\makebox(0,0){ 1}}
\color{black}
\color{black}
\thicklines \path(370,249)(1676,249)(1676,1418)(370,1418)(370,249)
\color{black}
\put(82,833){\makebox(0,0)[l]{\shortstack{$\mathcal{P(E)}$}}}
\color{black}
\put(1023,42){\makebox(0,0){$\mathcal{E}$}}
\color{black}
\put(749,424){\makebox(0,0)[l]{$\mathcal{E}_{\perp} = 0.4$}}
\color{grey}
\put(892,1137){\vector(0,-1){693}}
\color{magenta}
\thicklines \path(370,249)(370,249)(383,271)(396,294)(410,316)(423,339)(436,361)(449,384)(462,406)(476,428)(489,451)(502,473)(515,496)(528,518)(541,541)(555,563)(568,585)(581,608)(594,630)(607,653)(621,675)(634,698)(647,720)(660,742)(673,765)(687,787)(700,810)(713,832)(726,855)(739,877)(753,899)(766,922)(779,944)(792,967)(805,989)(819,1012)(832,1034)(845,1056)(858,1079)(871,1101)(884,1124)(898,1140)(911,1146)(924,1152)(937,1158)(950,1163)(964,1167)(977,1171)(990,1175)(1003,1178)(1016,1181)
\thicklines \path(1016,1181)(1030,1183)(1043,1185)(1056,1187)(1069,1188)(1082,1188)(1096,1188)(1109,1188)(1122,1187)(1135,1186)(1148,1184)(1162,1182)(1175,1179)(1188,1176)(1201,1172)(1214,1168)(1227,1163)(1241,1158)(1254,1153)(1267,1146)(1280,1140)(1293,1132)(1307,1124)(1320,1116)(1333,1107)(1346,1097)(1359,1087)(1373,1075)(1386,1064)(1399,1051)(1412,1038)(1425,1023)(1439,1008)(1452,992)(1465,975)(1478,957)(1491,938)(1505,918)(1518,896)(1531,872)(1544,847)(1557,820)(1570,791)(1584,760)(1597,725)(1610,686)(1623,642)(1636,592)(1650,530)(1663,449)(1676,249)
\color{black}
\thicklines \path(370,249)(1676,249)(1676,1418)(370,1418)(370,249)
\end{picture}

\caption{A Typical probability density for $\Pi_3$.  Note the point of discontinuity 
in the derivative at $\mathcal{E} = \mathcal{E}_{\perp}$}
\end{figure}
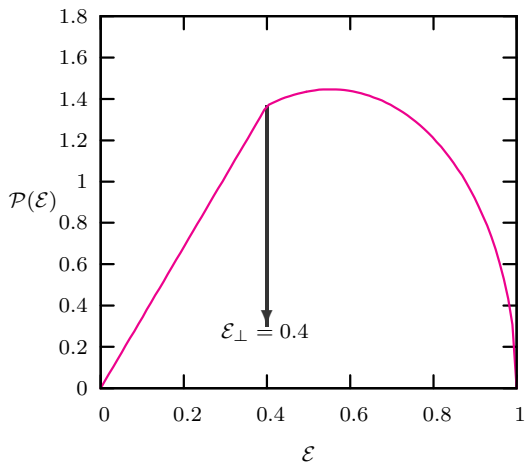

The integrating measure \cite{Byrd} may be conveniently written as $d\mathcal{H}_3= \sin 2\beta \sin 2\theta \sin^2 \theta d\alpha  d\beta  d\gamma  d\theta$, when the state is expanded in an orthonormal basis  as: $|\psi \rangle = \cos \theta |\chi_1 \rangle + e^{i (\alpha + \gamma)} \sin \theta \cos \beta |\chi_2 \rangle - e^{i (\alpha - \gamma)} \sin \theta \sin \beta |\chi_3 \rangle$, with the integration ranges, $\theta, \beta \in [0, \frac{\pi}{2}]$ and $\alpha, \gamma \in [0, \pi]$.
Conveniently, one may choose $\chi_{1,2}$ to be separable, and by a suitable LO, they can be
brought to the form $\vert \uparrow \uparrow \rangle,  | \downarrow\downarrow \rangle$.
We have verified that the resulting
probability density can be cast into  the simple form
\begin{eqnarray}
\mathcal{P}_3(\mathcal{E}) = \frac{2 \mathcal{E}}{\sqrt{1 - \mathcal{E}_{\perp}^2}} 
\cosh^{-1}{(\frac{1}{\mathcal{E}_{>}})}.
\end{eqnarray}
where $\mathcal{E}_{>} = \max (\mathcal{E}, \mathcal{E}_{\perp} )$.   

A typical curve for $\mathcal{P}_3(\mathcal{E})$ is shown in FIG. 2, which exhibits the required characteristic. 
The curve possesses a discontinuity in its derivative at 
$\mathcal{E}_{\perp}$. Significantly,
concurrence (being identically zero)  fails to distinguish different three dimensional projections, 
\textit{e.g.}, $\mathcal{E}_{\perp} =0 ~~\rm{or}~~ 1$, although their PDFs are vastly
different. 

Lastly, we consider the full space $\Pi_4$, whose PDF is universal.
This curve is obtained by using the Haar measure on $SU(4)$ \cite{Tilma:2002kf}. 
Note that the curve is
smooth everywhere, as shown in FIG. 4.
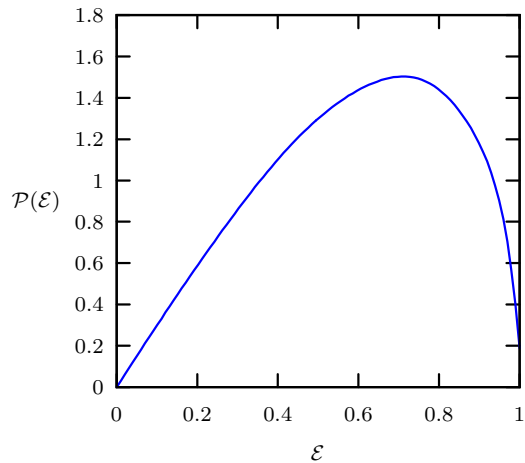
\begin{figure}[ht]

\setlength{\unitlength}{0.120450pt}
\begin{picture}(1800,1500)(0,0)
\footnotesize
\color{black}
\thicklines \path(411,249)(452,249)
\thicklines \path(1676,249)(1635,249)
\put(370,249){\makebox(0,0)[r]{$0$}}
\color{black}
\thicklines \path(411,379)(452,379)
\thicklines \path(1676,379)(1635,379)
\put(370,379){\makebox(0,0)[r]{$0.2$}}
\color{black}
\thicklines \path(411,509)(452,509)
\thicklines \path(1676,509)(1635,509)
\put(370,509){\makebox(0,0)[r]{$0.4$}}
\color{black}
\thicklines \path(411,639)(452,639)
\thicklines \path(1676,639)(1635,639)
\put(370,639){\makebox(0,0)[r]{$0.6$}}
\color{black}
\thicklines \path(411,769)(452,769)
\thicklines \path(1676,769)(1635,769)
\put(370,769){\makebox(0,0)[r]{$0.8$}}
\color{black}
\thicklines \path(411,898)(452,898)
\thicklines \path(1676,898)(1635,898)
\put(370,898){\makebox(0,0)[r]{$1$}}
\color{black}
\thicklines \path(411,1028)(452,1028)
\thicklines \path(1676,1028)(1635,1028)
\put(370,1028){\makebox(0,0)[r]{$1.2$}}
\color{black}
\thicklines \path(411,1158)(452,1158)
\thicklines \path(1676,1158)(1635,1158)
\put(370,1158){\makebox(0,0)[r]{$1.4$}}
\color{black}
\thicklines \path(411,1288)(452,1288)
\thicklines \path(1676,1288)(1635,1288)
\put(370,1288){\makebox(0,0)[r]{$1.6$}}
\color{black}
\thicklines \path(411,1418)(452,1418)
\thicklines \path(1676,1418)(1635,1418)
\put(370,1418){\makebox(0,0)[r]{$1.8$}}
\color{black}
\thicklines \path(411,249)(411,290)
\thicklines \path(411,1418)(411,1377)
\put(411,166){\makebox(0,0){$0$}}
\color{black}
\thicklines \path(664,249)(664,290)
\thicklines \path(664,1418)(664,1377)
\put(664,166){\makebox(0,0){$0.2$}}
\color{black}
\thicklines \path(917,249)(917,290)
\thicklines \path(917,1418)(917,1377)
\put(917,166){\makebox(0,0){$0.4$}}
\color{black}
\thicklines \path(1170,249)(1170,290)
\thicklines \path(1170,1418)(1170,1377)
\put(1170,166){\makebox(0,0){$0.6$}}
\color{black}
\thicklines \path(1423,249)(1423,290)
\thicklines \path(1423,1418)(1423,1377)
\put(1423,166){\makebox(0,0){$0.8$}}
\color{black}
\thicklines \path(1676,249)(1676,290)
\thicklines \path(1676,1418)(1676,1377)
\put(1676,166){\makebox(0,0){$1$}}
\color{black}
\color{black}
\thicklines \path(411,249)(1676,249)(1676,1418)(411,1418)(411,249)
\color{black}
\put(82,833){\makebox(0,0)[l]{\shortstack{$\mathcal{P(E)}$}}}
\color{black}
\put(1043,42){\makebox(0,0){$\mathcal{E}$}}
\color{blue}
\thicklines \path(411,249)(411,249)(424,269)(437,289)(449,308)(462,328)(475,348)(488,367)(500,387)(513,406)(526,426)(539,445)(552,464)(564,484)(577,503)(590,522)(603,541)(615,560)(628,578)(641,597)(654,616)(667,634)(679,652)(692,670)(705,689)(718,706)(730,724)(743,742)(756,759)(769,777)(782,794)(794,811)(807,828)(820,845)(833,861)(845,878)(858,894)(871,909)(884,925)(897,940)(909,955)(922,970)(935,985)(948,999)(960,1012)(973,1026)(986,1039)(999,1052)(1012,1064)(1024,1076)(1037,1087)
\thicklines \path(1037,1087)(1050,1098)(1063,1109)(1075,1119)(1088,1129)(1101,1139)(1114,1148)(1127,1156)(1139,1165)(1152,1172)(1165,1180)(1178,1187)(1190,1193)(1203,1199)(1216,1204)(1229,1209)(1242,1214)(1254,1217)(1267,1220)(1280,1223)(1293,1224)(1305,1225)(1318,1225)(1331,1224)(1344,1222)(1357,1219)(1369,1215)(1382,1209)(1395,1203)(1408,1195)(1420,1186)(1433,1175)(1446,1164)(1459,1151)(1472,1136)(1484,1120)(1497,1103)(1510,1084)(1523,1064)(1535,1041)(1548,1017)(1561,989)(1574,959)(1587,923)(1599,883)(1612,835)(1625,777)(1638,706)(1650,618)(1663,509)(1676,371)
\color{black}
\thicklines \path(411,249)(1676,249)(1676,1418)(411,1418)(411,249)
\end{picture}

\caption{The probability density $\mathcal{P}_4(\mathcal{E})$ for the entire Hilbert space.}
\end{figure}

It remains to consider the case when $\rho$ is an incoherent sum of the projections 
(see Eqn.3), where the weights have been chosen such that the special cases
$\rho = \Pi$ are naturally recovered; they also ensure that the results are 
not artefacts of any basis. If
$\vert \vert \rho_1 -\rho_2 \vert \vert$ is small, the corresponding 
probability density functions will also
be close to each other. 

\begin{figure}[ht]

\setlength{\unitlength}{0.120450pt}
\begin{picture}(1800,1500)(0,0)
\footnotesize
\color{black}
\thicklines \path(328,166)(369,166)
\thicklines \path(1676,166)(1635,166)
\put(287,166){\makebox(0,0)[r]{$0$}}
\color{black}
\thicklines \path(328,323)(369,323)
\thicklines \path(1676,323)(1635,323)
\put(287,323){\makebox(0,0)[r]{$0.2$}}
\color{black}
\thicklines \path(328,479)(369,479)
\thicklines \path(1676,479)(1635,479)
\put(287,479){\makebox(0,0)[r]{$0.4$}}
\color{black}
\thicklines \path(328,636)(369,636)
\thicklines \path(1676,636)(1635,636)
\put(287,636){\makebox(0,0)[r]{$0.6$}}
\color{black}
\thicklines \path(328,792)(369,792)
\thicklines \path(1676,792)(1635,792)
\put(287,792){\makebox(0,0)[r]{$0.8$}}
\color{black}
\thicklines \path(328,949)(369,949)
\thicklines \path(1676,949)(1635,949)
\put(287,949){\makebox(0,0)[r]{$1$}}
\color{black}
\thicklines \path(328,1105)(369,1105)
\thicklines \path(1676,1105)(1635,1105)
\put(287,1105){\makebox(0,0)[r]{$1.2$}}
\color{black}
\thicklines \path(328,1262)(369,1262)
\thicklines \path(1676,1262)(1635,1262)
\put(287,1262){\makebox(0,0)[r]{$1.4$}}
\color{black}
\thicklines \path(328,1418)(369,1418)
\thicklines \path(1676,1418)(1635,1418)
\put(287,1418){\makebox(0,0)[r]{$1.6$}}
\color{black}
\thicklines \path(328,166)(328,207)
\thicklines \path(328,1418)(328,1377)
\put(328,83){\makebox(0,0){$0$}}
\color{black}
\thicklines \path(598,166)(598,207)
\thicklines \path(598,1418)(598,1377)
\put(598,83){\makebox(0,0){$0.2$}}
\color{black}
\thicklines \path(867,166)(867,207)
\thicklines \path(867,1418)(867,1377)
\put(867,83){\makebox(0,0){$0.4$}}
\color{black}
\thicklines \path(1137,166)(1137,207)
\thicklines \path(1137,1418)(1137,1377)
\put(1137,83){\makebox(0,0){$0.6$}}
\color{black}
\thicklines \path(1406,166)(1406,207)
\thicklines \path(1406,1418)(1406,1377)
\put(1406,83){\makebox(0,0){$0.8$}}
\color{black}
\thicklines \path(1676,166)(1676,207)
\thicklines \path(1676,1418)(1676,1377)
\put(1676,83){\makebox(0,0){$1$}}
\color{black}
\color{black}
\thicklines \path(328,166)(1676,166)(1676,1418)(328,1418)(328,166)
\color{black}
\put(1110,859){\makebox(0,0)[l]{$\mathcal{E}_{\perp}$}}
\color{blue}
\thicklines
\path(449,297)(449,501)
\color{black}
\put(1137,1031){\vector(0,-1){146}}
\color{blue}
\thicklines 
\path(328,166)(328,166)(342,181)(355,195)(369,210)(382,225)(396,240)(410,254)(423,269)(437,284)(451,299)(464,313)(478,328)(491,343)(505,358)(519,372)(532,387)(546,402)(559,417)(573,431)(587,446)(600,461)(614,475)(628,490)(641,505)(655,519)(668,534)(682,549)(696,563)(709,578)(723,592)(736,607)(750,622)(764,636)(777,651)(791,665)(805,680)(818,694)(832,709)(845,723)(859,738)(873,752)(886,767)(900,781)(913,795)(927,810)(941,824)(954,838)(968,852)(982,867)(995,881)(995,881)(1009,895)(1022,909)(1036,924)(1050,938)(1063,952)(1077,967)(1091,981)(1104,996)(1118,1010)
\thicklines \path(1131,1025)(1145,1035)(1159,1041)(1172,1047)(1186,1054)(1199,1060)(1213,1066)(1227,1073)(1240,1080)(1254,1087)(1268,1095)(1281,1104)(1295,1114)(1308,1125)(1322,1138)(1336,1153)(1349,1173)(1363,1198)(1376,1233)(1390,1285)(1404,1370)(1417,1310)(1431,1237)(1445,1182)(1458,1140)(1472,1107)(1485,1078)(1499,1053)(1513,1030)(1526,1008)(1526,686)(1553,665)(1567,642)(1581,616)(1594,587)(1608,554)(1622,515)(1635,469)(1649,413)(1662,341)(1676,203)
\color{black}
\thicklines \path(328,166)(1676,166)(1676,1418)(328,1418)(328,166)
\end{picture}

\caption{The overall probability density $\mathcal{P}_4(\mathcal{E})$ for a typical mixed state, $\rho$, with eigenvalues \{0.385, 0.288, 0.231, 0.096\}. Note that the features of
the indivdual subspaces are vividly preserved.}
\end{figure}
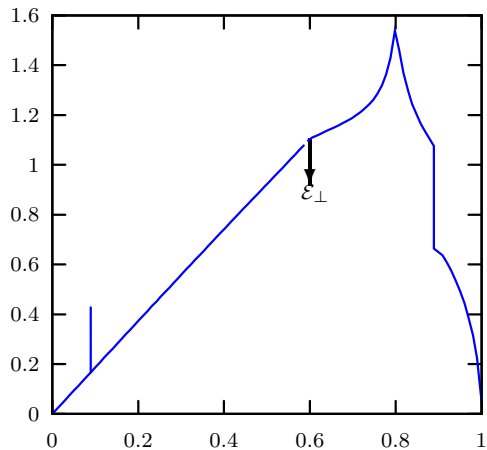

FIG. 4 illustrates the PDF for this general case. The important
point to be noted is that the superposition of curves \textit{does not} obliterate 
the information
contained in individual curves; they are retained as points of discontinuity 
or singularity and each
individual PDF may be reconstructed, together with the associated weights.
$\mathcal{P(E)}$ is by definition invariant under LO. With this,
one may ask if the state itself may be reconstructed, up to LO. Before we take up 
this question,
we consider an important application of this prescription, to NMR QC.

NMR QC employs the so called pseudopure states for computation. Since 
it is experimentally demonstrated that  the quantum logic  operations used in QC are
 implementable with NMR,
it follows that  these states should possess a non vanishing entanglement. Indeed, they have the form $\rho_{pps} = \frac{1 - \epsilon}{4} \mathbf{1} + \epsilon |\psi \rangle \langle \psi |$, in our system of expansion. The NMR signal is sensitive only to the pure component, the so called deviation matrix. Accordingly, its $\mathcal{P}_{\rho}(\mathcal{E})$ is given by a weighted
Dirac Delta superposed on the PDF  coming from the full space. 
The uniform background
is invariant under unitary operations, but the one dimensional fluctuation is not, 
allowing
for non-trivial gate operations. Thus NMR QC exploits the excess of entangled states over the unpolarized background as a resource, and this feature is correctly captured by the 
PDF of the state. This is in contrast to other measures which attribute a zero entanglement to all
PPS with $ \epsilon \leq \frac{1}{3}$, while usually in practice in NMR QC $\epsilon \sim 10^{-6}$ . This analysis also raises the interesting possibility of QC with more general
pseudo projection states.\\

Lastly, we return to the issue of the reconstructibility of the state (up to LO).
If $\rho$ is a projection, the reconstructibility is assured, by construction. When $\rho$ is
more general, the reconstruction is partial. For, the action of $SU(2) \times SU(2)$ on
$\rho$ produces an orbit of dimension six, characterised by nine invariants. The set of parameters
which characterize the entanglement are seven in number (for example: $\{\mu_1, \mu_2, \mu_3, \mathcal{E}_{1}, \mathcal{E}_{cusp}, \mathcal{E}_{max}, \mathcal{E}_{\perp} \}$ ). Geometrically, 
$\mathcal{P}_{\rho}(\mathcal{E})$ is invariant under independent LO, $L_i$, acting on 
the subspaces
$\Pi_i$, where $\Pi_i \subset \Pi_{i+1}$. If $\rho$ is to be 
unique up to a global LO, one needs the additional constraint $L_i = U L_{i}^{(0)}$,
where $L_{i}^{(0)}$ may be chosen freely. Let us choose $L_2^{(0)} = \mathbf{1}$ 
(where $\mathbf{1}$ is the identity operator).  The nestedness condition, \textit{viz.}, that $|\psi_1\rangle \in \Pi_2$ and $|\psi_4\rangle \in \Pi_2^c$ , entails that $L_1^{(0)}$ and  $L_3^{(0)}$ get specified by two parameters each
\footnote{In fact we need only one parameter each to fix the states up to discrete 
ambiguities.  This is because from the $\mathcal{P}_{i}(\mathcal{E})$ we already 
 know the entanglement of the states, which fixes one parameter.  However, 
 there remains a discrete ambiguity if only one of $\theta$ or $\phi$ is specified}.

  More explicitly, if we have $\Pi_2$ in the canonical form, it is 
  spanned by $|\chi_1 \rangle$ and $|\chi_2\rangle$ 
  given respectively as: $(1,0,0,0)$ and $(0,x,y,z)$.  Therefore, 
  we can specify $|\psi_1\rangle = 
  |\chi_1 \rangle \cos{\frac{\theta}{2}} e^{i\phi / 2} + 
  |\chi_2 \rangle \sin{\frac{\theta}{2}} e^{- i\phi / 2}$ 
  by giving the values of $(\theta, \phi)$. Similarly, $|\psi_{\perp} \rangle$ 
  can be specified by $(\theta_{\perp}, \phi_{\perp})$ when it is expanded 
  in the canonical basis of $\Pi_2^{c} = 
  (\mathbf{1} - \Pi_2)$, given by $|\chi_1^{c} 
  \rangle = (0,0,c/ \sqrt{c^2+b^2},-b/ \sqrt{c^2+b^2})$ and 
  $|\chi_2^{c} \rangle = (0, \sqrt{c^2+b^2}, ab/ \sqrt{c^2+b^2}, 
  ac/ \sqrt{c^2+b^2})$.\\

In conclusion, we have given a  prescription that describes the entanglement of mixed states by not just a number, but an exhaustive set of parameters which characterize the manner in which the entanglenment is distributed over the ensemble. They further permit an almost complete reconstruction of the state up to LO.  The prescription may   provide a better insight into  other measures of entanglement such as entanglement of distillation and  entanglement cost.  Investigations along these lines,  and a further study of the PDFs  for higher spins may provide us with a better appreciation of quantum entanglement.

\end{document}